\documentclass[twocolumn,secnumarabic,amssymb, nobibnotes,superscriptaddress,showpacs, prb]{revtex4-1}
\usepackage{bm}
\usepackage{color}
\usepackage{graphicx}
\usepackage{multirow}
\usepackage{ulem}
\usepackage{subfigure}
\usepackage{amsmath}
\usepackage{xspace} 
 \usepackage{adjustbox}
\usepackage[utf8]{inputenc}

\newcommand{\STO}{{SrTiO$_3$}\xspace}
\newcommand{\BTO}{{BaTiO$_3$}\xspace}

\begin{document}
\title{\textit{Ab initio} based study on atomic ordering in {(Ba, Sr)}TiO$_3$}
\author{Aris Dimou}
\affiliation{Interdisciplinary Centre for Advanced Materials Simulation (ICAMS) and Center for Interface-Dominated High Performance Materials (ZGH), Ruhr-University Bochum, Universit\"atsstr 150, 44801 Bochum, German}
\email{aris.dimou@rub.de}
\author{Ankita Biswas}
\affiliation{Interdisciplinary Centre for Advanced Materials Simulation (ICAMS) and Center for Interface-Dominated High Performance Materials (ZGH), Ruhr-University Bochum, Universit\"atsstr 150, 44801 Bochum, German}
\affiliation{Department of Materials Science and Engineering, University of Virginia, Charlottesville, Virginia 22902, United States}
\author{Anna Gr\"unebohm}
\email{anna.gruenebohm@rub.de}
\affiliation{Interdisciplinary Centre for Advanced Materials Simulation (ICAMS) and Center for Interface-Dominated High Performance Materials (ZGH), Ruhr-University Bochum, Universit\"atsstr 150, 44801 Bochum, German}
\date{\today}

\begin{abstract}

\ We combine density functional theory and molecular dynamics simulations to investigate the impact of Sr concentration and atomic ordering on the structural and ferroelectric properties of (Ba, Sr)TiO$_3$. On one hand, the macroscopic structural properties are rather insensitive to atomic ordering. On the other hand, the Curie temperature and polarization differ by $9$\% and $17$\% for different symmetries of the Sr distribution, respectively. Local ordering of Sr induces preferential polarization directions and influences the relative stability of the three ferroelectric phases.

\end{abstract}

\maketitle
\section{Introduction}
\begin{figure*}[t]
    \centerline
    {\includegraphics[height=0.365\textwidth,clip,trim=4cm 10.7cm 5cm 6cm]{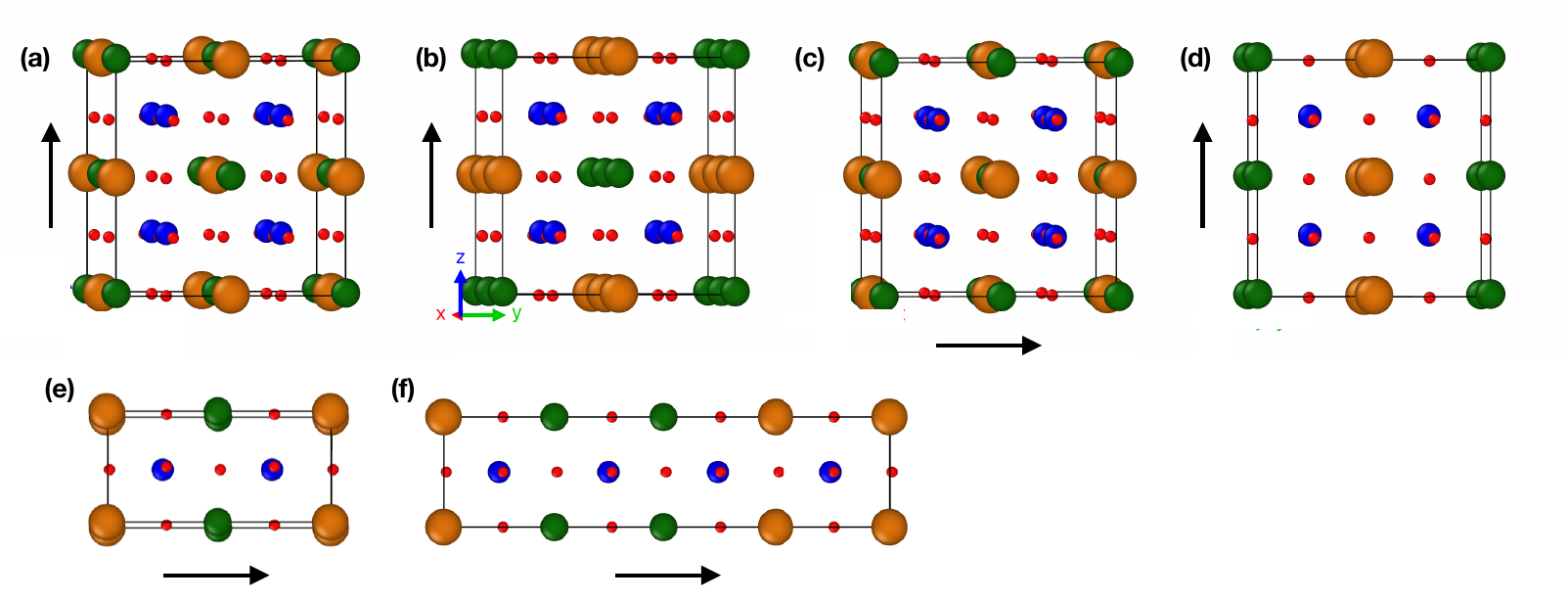}}
    \caption{Illustrations of the considered ordered structures (OS) (a) 0-d (rock-salt ordering), (b)  1-d$_{\perp}$ (columns of Sr perpendicular to tetragonal axis), (c) 1-d$_{||}$ (columns of Sr parallel to tetragonal axis), (d) 2-d$_{||}$ (planes of Sr parallel to tetragonal axis), (e) 
    2-d$_{\perp}$ or 2-d$_{||}$ (planes of Sr perpendicular to the tetragonal axis), and (f) double-layer perpendicular to the tetragonal axes. With Sr, Ba, Ti, and O represented in green, orange, blue, and red. The black arrow indicates the polarization direction of the corresponding supercell.
    }
\label{fig:Intro_Struc_atom_ordering}
\end{figure*}

\ Lead-free ferroelectric solid solutions based on  BaTiO$_3$ are used in a wide range of applications due to their dielectric, piezoelectric, pyroelectric, and caloric functional responses\cite{acosta_batio_2017, moya_caloric_2014, said_ferroelectrics_2017, lin_excellent_2019} and phase transition temperatures and functional properties are commonly optimized by substitution.\cite{acosta_batio_2017, bai_enhanced_2018, sun_defect_2021} In particular solid solutions of BaTiO$_3$ and SrTiO$_3$ are of high technological importance.\cite{acosta_batio_2017} Already pristine \BTO exhibits a complex phase diagram with the paraelectric cubic (C) phase at high temperatures and transitions to the tetragonal (T), orthorhombic (O) and rhombohedral (R) phases at $398$~K, $273$~K and  $183$~K, respectively.\cite{lines_principles_1977} Replacing Ba by smaller Sr ions systematically reduces transition temperatures, strain, polarization, and volume\cite{bunting_properties_1947, durst_solid_1950, lemanov_phase_1996, menoret_structural_2002, acosta_batio_2017,nishimatsu_molecular_2016,mcquarrie_structural_1955, menoret_structural_2002, bunting_properties_1947, kim_origins_2014} while SrTiO$_3$ is considered a quantum paraelectric material\cite{zhang_cell_1997, guzhva_ferroelectric_1998} with a transition between cubic and tetragonal paraelectric phases at  $110$~K.\cite{sai_first-principles_2000,shirane_lattice-dynamical_1969} As changing the concentration of Sr by $20$\% reduces the temperature of the  C to T transition by $50$~K,\cite{lemanov_phase_1996} it has to be expected that local variations of the concentration present in most samples\cite{fuks_ab_2005} impact local polarization and the phase transitions. However, this aspect is so far underrepresented in literature. 

\ Recently, targeted inhomogeneities in nanostructures, particularly superlattices, became an important playground for the optimization of ferroelectric phases and functional properties.\cite{grunebohm_interplay_2022} For instance vortex states, rotational polarization, negative capacitance, and double hysteresis loops have been reported for these inhomogeneous materials.\cite{hong_vortex_2009, estandia_rotational_2019, peng_three-dimensional_2019,wang_direct_2020,aramberri_ferroelectricparaelectric_2022} For nanostructures based on \BTO{} and \STO ferroelectric polarization in SrTiO$_3$ layers,\cite{neaton_theory_2003, lee_origin_2009} original domain patterns \cite{lisenkov_phase_2007} as well as pinning of domain walls and unusual domain evolution have been reported.\cite{stepkova_pinning_2018, dimou_pinning_2022}  However, there are important gaps in knowledge on the impact of atomic ordering at these intrinsic interfaces on the material properties. On one hand, it is difficult to explain these experimentally. On the other hand, the details of the atomic ordering are so far neglected in large-scale simulations like the effective Hamiltonian approach.\cite{walizer_finite-temperature_2006, nishimatsu_molecular_2016}

\ Atomistic simulations successfully reproduce the phase diagram of (Ba,Sr)TiO$_3$ solid solutions with random ordering and it has been reported that the local polarization increases in the vicinity of Sr dopants due to the smaller ionic radii.\cite{tinte_ferroelectric_2004,wexler_sr-induced_2019,dimou_pinning_2022} While this observation was restricted to the rhombohedral phase for random ordering\cite{wexler_sr-induced_2019} and the increase has also been observed for the tetragonal phase in case of layered ordering.\cite{dimou_pinning_2022} For the tetragonal phase specific atomic orderings for 25\% Sr were compared in Ref.~\onlinecite{rusevich_electromechanical_2017} utilizing density functional theory at $0$~K. It has been discussed that the ordering may reduce the symmetry and induce a change of the piezoelectric coefficient $d_{33}$ of about $7$\%. While such differences in the piezoelectric response may be insignificant, a corresponding spread of the local polarization may significantly impact functional properties at finite temperatures. Furthermore, to the best of our knowledge, neither the dependency of Born charges nor of high-frequency dielectric constant on atomic ordering have been reported. These are, however, important input parameters for coarse-grained simulation methods\cite{walizer_finite-temperature_2006} and the fundamental understanding of the ferroelectric instability.\cite{ghosez_born_1995, cohen_origin_1992}

\ In this paper, we thus study the impact of Sr concentration and ordering on ferroelectric phase diagrams and functional properties by molecular dynamics simulations. For the tetragonal phase, we furthermore use density functional theory to compare atomic orderings with specific symmetries as transition between random ordering and ordered nanostructures. We find that for one fixed concentration, atomic ordering may induce a spread of Curie temperature and polarization of $9$\% 
$17$\%.

\begin{figure}[t]
    \centering
        \includegraphics[width=0.4\textwidth,clip,trim=11.9cm 3cm 11.9cm 3cm] {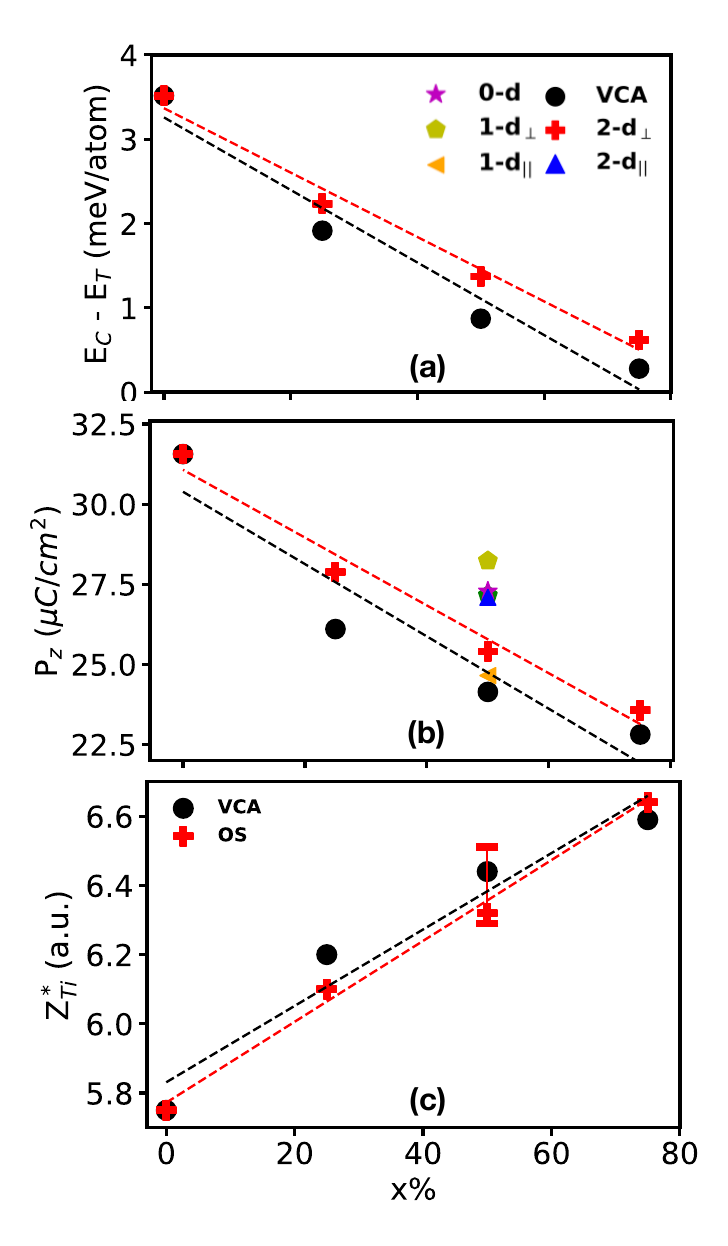}
    \caption{Change of properties of the tetragonal phase of BaTiO$_3$ with Sr substitution comparing the ordered structures (2-d$_{\perp}$, colored symbols) depicted in Fig.~\ref{fig:Intro_Struc_atom_ordering} to the homogeneous distribution approximate with the virtual crystal approximation (VCA, black dots). (a) The energy difference between the cubic paraelectric ($E_C$) and the tetragonal ferroelectric phase ($E_T$), (b) Macroscopic polarization $P_z$ (the double layer is also at $27.2$~$\mu$C/cm$^2$), and (c) mean values of Born effective charges $Z^{*}_{33}$. Dashed lines show the mean square fit.}
    \label{fig:Static_IAO}
\end{figure}

\section{Computational Details}

\ First-principles, density functional theory (DFT), and density functional perturbation theory\cite{grling_dft_1995} simulations are performed with the Abinit package.\cite{gonze_recent_2016} The exchange-correlation energy is approximated with the generalized gradient approximation (GGA). Precisely we use the GGA modification proposed by Wu and Cohen.\cite{wu_more_2006} We use a $8\times 8 \times 8$ Monkhorst-pack k-mesh,\cite{monkhorst_special_1976} cutoff energy of at least $63$~Hartree together with energy and force thresholds for self-consistent cycles and structural optimization of $10^{-8}$ Hartree and $10^{-6}$ Hartree/Bohr, respectively.

\ We focus on the tetragonal phase and compare the properties of homogeneous, randomly ordered systems approximated by the virtual crystal approximation\cite{bellaiche_virtual_2000}  with chosen ordered structures (OS) in super-cells with up to $40$ atoms. For $50$\% Sr, we consider the six structures shown in the subfigures of Figure~\ref{fig:Intro_Struc_atom_ordering}: (a) The rock-salt structure (0-d); Columns of Sr (b) perpendicular (1-d$_{\perp}$) and (c) parallel (1-d$_{||}$) to the tetragonal direction; Layered 2-d configurations (d) parallel (2-d$_{||}$) or (e) perpendicular (2-d$_{\perp}$,) to the tetragonal axes. The latter case corresponds to a superlattice with a periodicity of one, i.e., a mono-layer and we also include a double-layer shown in (f) with a periodicity of two in our comparison. Only the rock-salt structure has the same symmetry as the randomly ordered solid solution, while already in the paraelectric phase, the symmetry of the material is reduced for 1-d and 2-d orderings. Note that for 1-d$_{\perp}$ and 2-d$_{\perp}$ the symmetry of the ferroelectric phase is reduced to monoclinic. For $25$\% and $75$\%, we use a $1\times 1\times 4$ cell with 2-d$_{\perp}$ symmetry which is energetically most favourable for $50$\%.

\ We calculate the local polarization as the dipole moment per Ti-centered unit cell as
\begin{equation}
    \label{eq:polarization}
    \vec{P}=\frac{1}{V}\sum\limits_{i}\frac{1}{w_i}Z_{i}^*\vec{r_i}\,, 
\end{equation}
where $V$ is the volume of the unit cell, $w_i$ is a weight factor corresponding to the number of unit cells that share atom $i$, and $Z_i$ is the Born effective charge of an atom $i$. Finally, $r_i$ is the displacement of atom $i$ with respect to the centrosymmetric cubic state. 

\ Supplementary, we utilize the core-shell potentials parametrized by Sepliarsky and Tinte\cite{sepliarsky_atomic-level_2005,tinte_ferroelectric_2004} to investigate the influence of finite temperature and thermal fluctuation on ferroelectric polarization and phase transition temperatures for  0-d, random, and 2-d${\perp}$ configurations. This model\cite{mitchell_shell_1993} allows for polarizability by splitting each atom into a positively charged core and a negatively charged shell that is linked to its core via a spring. Furthermore, interactions between ions include long-range Coulomb and short-range Buckingham potentials, with a cut-off of $12$~\AA{}.  

\ We perform molecular dynamics simulations (MD) using LAMMPS,\cite{thompson_lammps_2022} where the Coulomb interactions are computed using the particle-particle particle-mesh method. We use a system size of $16$$\times$$16$$\times$$16$ with $40,960$~atoms. The MD simulations are conducted in the NPT ensemble, with temperature and pressure being controlled via a Nos\'e-Hoover thermostat and barostat with relaxation times of $0.04$ and $0.1$~ps, respectivley. Using a time-step of $0.4$~fs we record thermal averages for $20$~ps after equilibration of $5$~ps, in the case of the local polarization we recorded thermal averages for $100$~ps. For the layered configuration we furthermore sample the field response to an electric field applied perpendicular to the polarization direction. Therefore we model field ramping by changing the electric field in increments of $\pm 40$~kV/cm at $120$~K, i.e., in the tetragonal phase.

\section{Results}
\begin{figure}[t!]
    \centering
    \includegraphics[width=0.5\textwidth,clip,trim=6cm 4cm 4cm 3cm] {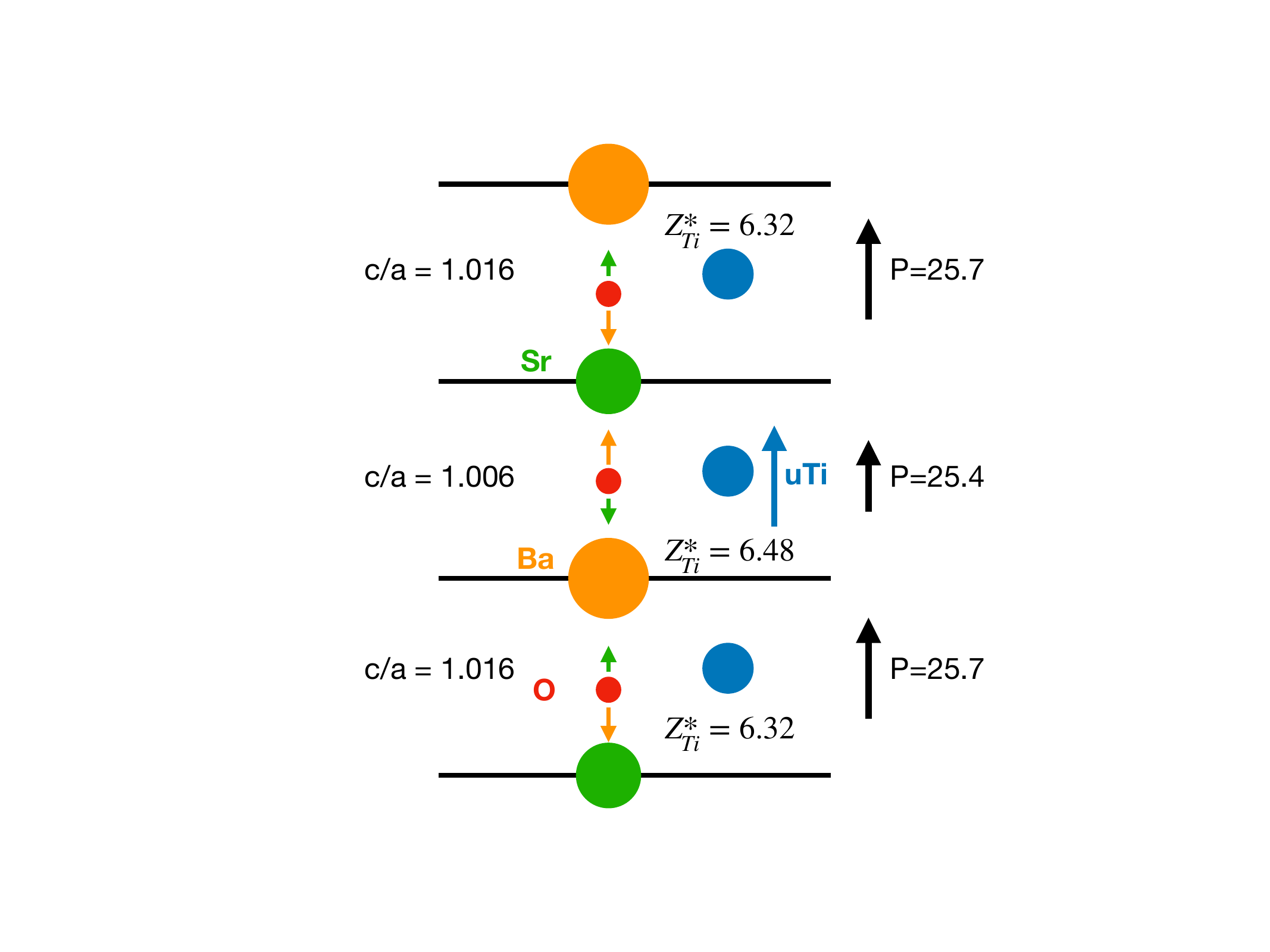}
    \caption{Illustration of the ``layered'' changes of polarization and tetragonal distortion if the polarization points perpendicular to the stacking. The red spheres represent oxygen atoms, while the orange, green, and blue spheres represent Ba, Sr, and Ti atoms respectively. The arrows in the colors of the corresponding atoms indicate the force vectors of the Ba-O and Sr-O repulsion (schematic representation), while the blue arrow represents the displacement vector of Ti ($u_{Ti}$). Due to the less repulsive Sr-O interaction, the O atom may shift more against the Ti atom, which can favor the polarization direction. However, the opposite effect can also occur, resulting in a lower polarization. The Born effective charges of the corresponding unit cell are indicated.}
    \label{fig:A-O_repuslion}
\end{figure}

\subsection{Static properties at $T=0$~K}

\ We start with the DFT study on the impact of the Sr concentration on the tetragonal ferroelectric phase for {2-d$_{\perp}$} ordering and VCA while for $50$\% Sr we consider all atomic orderings shown in Fig.~\ref{fig:Intro_Struc_atom_ordering}. Generally, both the different atomic orderings and the virtual crystal approximation reproduce the systematic reductions of volume and ferroelectric properties with Sr substitution as reported in the literature. Exemplary, Figures~\ref{fig:Static_IAO}~(a) and (b) show the reduction of the energy difference between ferroelectric and paraelectric state and a reduction of the macroscopic polarization. The reduction of volume and tetragonal distortion as well as the increase of the high-frequency dielectric constant are summarized in the appendix (Fig.~\ref{fig:appendix1}). 

\ However the polarizability of the Ti-O bonds and  ${\epsilon}^{\infty}_{33}$ show the opposite trend and increase with the Sr-concentration. The former is illustrated by the linear increase of the Born charges of the Ti-atoms and is in agreement with the reported larger polarizability of STO,\cite{furuta_first-principles_2010} see  Fig.~\ref{fig:Static_IAO}~(c). footnote{Note that due to the charge neutrality condition, the corresponding charges of the O-atoms are -$4.6e$ for the pristine material and show a slope of $-0.0117e$/Sr concentration(\%).} Note that the Born charges on the A-sites and of O-atoms perpendicular to the Ti-O bond are close to their nominal charge of $\pm 2e$ for all concentrations (for O: $-1.95 e$, slope:$-0.0006e$/Sr) see Fig.~\ref{fig:appendix1}. Moreover, only the Z$^*$ of the polar axis is shown. Quantitatively, the spreads of volume and ${\epsilon}^{\infty}$ of the different configurations at $50$\% Sr, is below $1$\% and $3$\%, respectively see Table~\ref{tab:BSTO_o}. Although the Born effective charges measure changes in the electronic structure under structural distortions and are thus potentially most sensitive to atomic ordering, the differences between VCA and the mean values for different atomic orderings are below $3$\%. Furthermore, the differences in local Born charges are small for most orderings. Only in the case of 2d$_{||}$ ordering, i.e., for 2-d$_{\perp}$ and double-layers, the dynamic charge transfer along the Ti-O bond differs between the Ti-positions, see Fig.~\ref{fig:A-O_repuslion}. In the case of the double-layer the difference between Z$_{Ti}^*$ in pure Sr and Ba cells reaches $10$\%. With the former being larger.

\ The polarization is most sensitive to ordering with variations of its macroscopic value of up to $17$\%. The maximal and minimal values are found for {1-$d_{\perp}$} and {1-$d_{||}$} configurations, respectively. For 2-d ordering, it is the opposite way around with larger polarization for the $d_{||}$ configuration. In both cases, the configurations with the largest polarization also show the largest tetragonal strain. For the 0-d ordering the polarization is similar to 2-d$_{||}$ and the macroscopic polarization for 2-d$_{\perp}$ increases with inhomogeneity, i.e., between mono- (2-d$_{\perp}$) and double-layers. The latter has been related to the modification of the local soft mode at the interface of BaTiO$_3$/SrTiO$_3$.\cite{lee_origin_2009} Importantly, VCA by construction cannot reproduce these differences and results in a smaller polarization than the discussed orderings. Furthermore, the symmetry breaking for 1-d$_{\perp}$ and 2-$d_{||}$ structures, i.e., the alternating Sr-Ba rows/planes along one direction perpendicular to the polarization, induces a monoclinic distortion and non-zero local polarization components along that direction of $\pm 1.2$~$\mu$ C/cm$^2$ and $\pm 2.5$~$\mu$ C/cm$^2$, correspondingly. However, note that the signs of these local polarization components also alternate resulting in zero net polarization.

\ Analog to the Born charges, also local strain and polarization show significant variations for $2-d_{\perp}$ ordering, only. In the case of the monolayer, both strain and polarization are about $1$\% larger in cells with polarization pointing towards Ba. Figure~\ref{fig:A-O_repuslion} explains this ``layered'' polarization: First, Born charges are slightly enhanced if the polarization points towards the Ba-rich cells. Furthermore, in this case, the oxygen atoms,  sketched in red, shift towards Sr. If the polarization points towards Sr, the oxygen atoms shift towards Ba.   Since the Sr-O interaction is less repulsive,\cite{furuta_first-principles_2010} the oxygen shift is considerably larger in the former case resulting in a slightly larger polarization (cf.\ Eq.~\eqref{eq:polarization}). Analogous, in the double-layer,  polarization (and strain) at the Ba-Sr interface accounts for $27.33$~$\mu$C/cm$^2$  ($1.021$ ) and $26.78$~$\mu$C/cm$^2$  ($1.004$) if the polarization points toward Ba and Sr, respectively. In this configuration, furthermore pure Ba units with $28.11$$\mu$C/cm$^2$ and pure Sr units with $26.40$$\mu$C/cm$^2$ exit. Note that for this inhomogeneous ordering the tetragonal strain reaches $1.035$, while the pure Sr cells are compressed to $0.993$.\footnote{Note that due to the depolarization at the \BTO, \STO interface, there is no quantitative correlation between strain and polarization.}

\ For the 2-d$_{\perp}$ structures, we also analyze the impact of atomic ordering on the paraelectric phase. While the macroscopic polarization is always zero, the symmetry reduction induces persistent tetragonal strain. As the oxygen atoms on Ba-O-Sr rows relax towards Sr, furthermore alternating polarization of $\pm 0.5~\mu$C/cm$^2$, for the 2-d$_{||}$ and  $\pm 0.8$~$\mu$C/cm$^2$ for the double-layer with the unit cells at the BaTiO$_3$/SrTiO$_3$ acting as a buffer, persist.

\ In summary, atomic ordering changes the local symmetry and thus affects the (local) polarization. However, one has to ask if the discussed orderings are possible in real samples. For this purpose, we compare the energies of their tetragonal phases, given in Table~\ref{tab:BSTO_o}. The energy differences between the tetragonal structures correlate with the symmetry of the configurations. The symmetric 0-d structure is about $6$~meV/atom higher in energy than the most favorable least symmetric 2-d structures. This spread of energies would correspond to a thermal energy of $68$~K only (using $E\approx k_BT$). Assuming a similar energy sequence in the cubic phase all these orderings may locally form during synthesis. Furthermore, once a certain atomic ordering has locally formed during synthesis, it breaks the symmetry and may favor either polarization perpendicular to the Sr-columns (1-d case) or parallel to the Sr-planes (2-d case). However, the latter energy difference is approaching the limit of our accuracy.

\begin{table}[t]
    \centering
    \caption{Impact of atomic ordering on macroscopic lattice constants ($a,b,c$), polarization along the tetragonal axis ($P_z$), dielectric constant $\epsilon^{\infty}_{33}$ for tetragonal axis Ba$_{50}$Sr$_{50}$TiO$_3$ and energy difference to the 2-d$_{\perp}$ ($\Delta E$)}
    \label{tab:BSTO_o}
    \begin{tabular}{c|ccccc}
    \hline
          & a & c/a & P$_z$ & {$\epsilon^{\infty}_{33}$} & $\Delta E$  \\
              &(\AA) &  & ({$\mu$C/cm$^2$})&  & (meV/atom) \\
\hline
    VCA            & 3.940 & 1.011 & 24.1 & 6.05 & -  \\
    0-d~~          & 3.938 & 1.012 & 27.3 & 5.96 & 6.2\\ 
    1-d$_{||}   $  & 3.940 & 1.012 & 24.7 & 6.07 & 3.8\\
    1-d$_{\perp}$  & 3.931 & 1.014 & 28.2 & 5.93 & 3.2\\
    2-d$_{||}$     & 3.934 & 1.013 & 27.1 & 5.96 & 0.0\\ 
 2-d$_{\perp}$     & 3.937 & 1.011 & 25.4 & 6.00 & 0.1\\ 
    double-layer   & 3.935 & 1.013 & 27.2 & 5.94 & 0.9\\
\hline
    \end{tabular}

\end{table}

\begin{figure*}[t!]
  \centerline{\includegraphics[height=0.65\textwidth,clip,trim=.75cm 3.5cm 1cm 0.25cm]{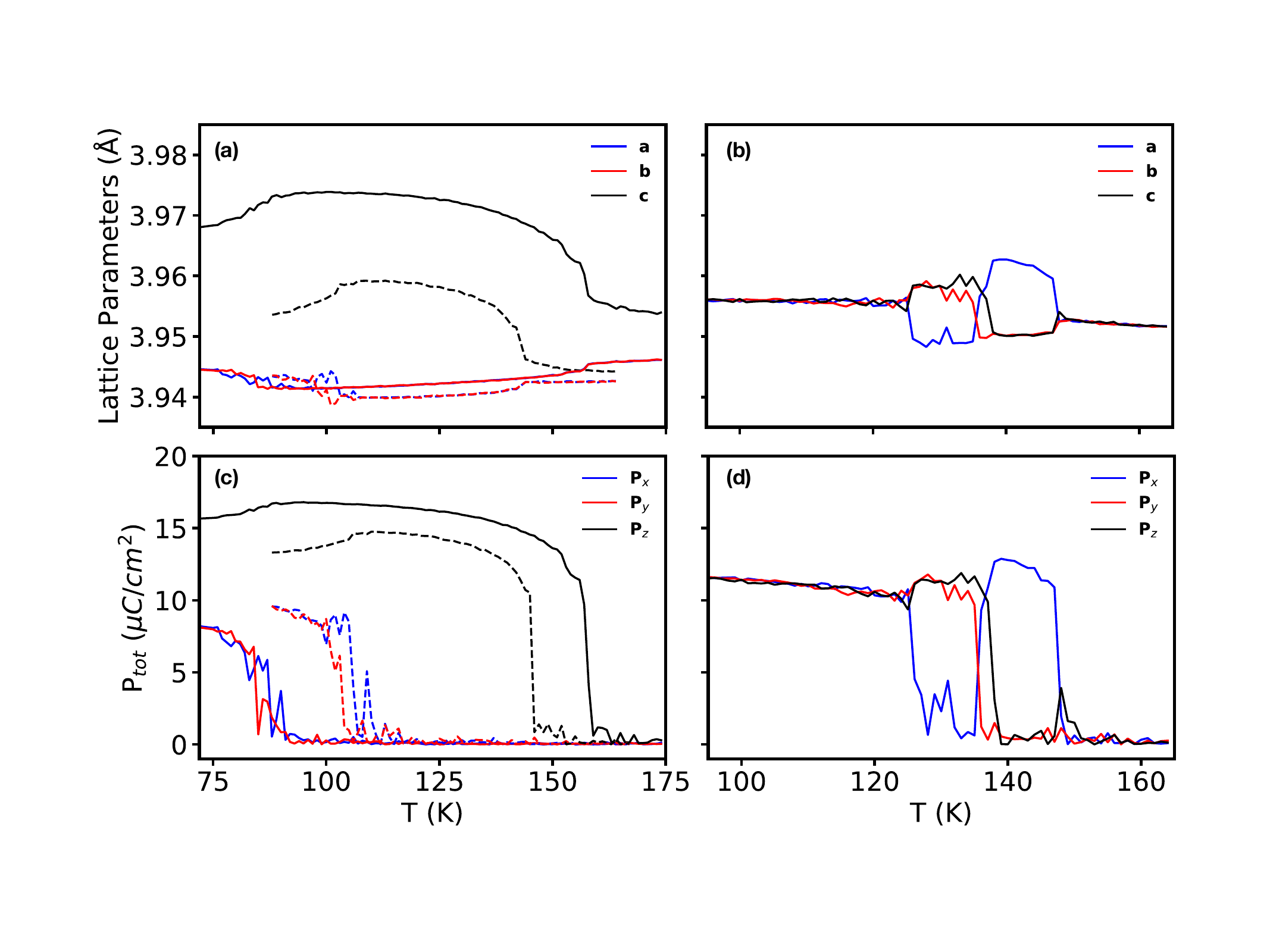}}
  \caption{Change of (a)--(b) lattice parameters and (c)--(d) polarization components found in heating simulations for (a)/(c) 2-d$_{\perp}$ and double-layer ordering and (b)/(d) 0-d rock-salt ordering. Colors highlight the three cartesian directions and in (a)/(c) results for 2-d$_{\perp}$ and double-layers are shown with dashed and solid lines, respectively.
  }
\label{fig:RS_Tc}
\end{figure*}

\begin{figure}[t]
  {\includegraphics[width=0.45\textwidth,clip,trim=0cm 0cm 0cm 0cm]{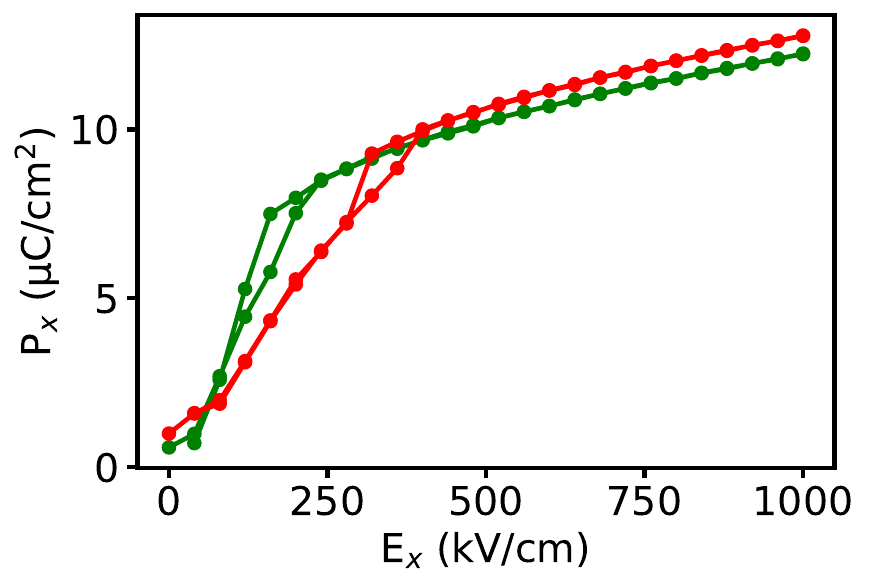}}
  \caption{Field-induced polarization rotation of 2-d$_{\perp}$ (green) and double-layer (red) structure with initial polarization along $z$ at 120~K for an electric field applied along $x$. Note that the symmetric hysteresis has to be expected for negative fields.}
\label{fig:PE_DB_TC}
\end{figure}

\subsection{Phase diagrams at finite temperature}

\ Do ordered and random configurations show the same trends with temperature? To answer this question we additionally perform molecular dynamics simulations with empirical potentials and record the phase sequence under heating for rock-salt, 2-d$_{||}$, and double-layer structures of Ba$_{50}$Sr$_{50}$TiO$_3$, see Fig.~\ref{fig:RS_Tc}, as well as for five independent random configurations shown in the appendix (cf.\ Fig.~\ref{fig:Samples_stat}). 

\ Indeed the randomly ordered structures and the high symmetric rock-salt structure all show the expected phase sequence from rhombohedral to orthorhombic, to tetragonal, and finally to cubic paraelectric phases with transition temperatures of $126$~K, $136$~K, and $148$~K for the latter and T$_c$ of $144\pm2$~K for the random configurations, see Fig.~\ref{fig:RS_Tc}. Experimental values between~$245-295$~K  have been reported.\cite{lemanov_perovskite_1999, menoret_structural_2002, kim_electrical_2013} Note that the underestimation of $T_c$ is the expected trend for DFT derived empirical potentials.\cite{tinte_ferroelectric_2004}

\ On the other hand, the reduced symmetry in the layered structures strongly modifies the phase diagram. Particularly, a lattice expansion along the stacking normal, i.e., along $z$ is observed for all temperatures.~\footnote{Note that core-shell potentials tend to quantitatively overestimate the tetragonal distortion, e.g., for pristine BTO $T=0$~K c/a$=1.029$ compared to a DFT value of $1.023$.} This strain along $z$ results in a monoclinic distortion of the low-temperature phase with $c>a=b$. In the tetragonal phase strain and polarization increase to $1.008$ ($1.005$), $16.4$~$\mu$C/cm$^2$  ($13.7$~$\mu$C/cm$^2$), at $120$~K for the double-layer (2-d$_{||}$, mono-layer) compared to $1.003$ and $11.4$~$\mu$C/cm$^2$ in the rock-salt structure at $145$~K. Furthermore, the high-temperature phase at $170$~K is indeed only pseudocubic with a persisting tetragonal distortion perpendicular to the Sr-layers of $1.002$ ($1.001$) and alternating local polarization of $\pm1.5$~{$\mu$C/cm$^2$} for double (and 2-d$_{||}$) structures. The atomic ordering furthermore stabilizes the tetragonal phase between $89$~K and $157$~K ($104-143$~K). Thus, the lower $T_c$ of the layered structures is reduced below the O-R transition temperature for rock-salt ordering and it is not clearly possible to distinguish the orthorhombic phase from thermal noise, cf.~Figure~\ref{fig:RS_Tc}.\footnote{Note that although the energy difference between tetragonal and orthorhombic phases of the rocksalt structure of $0.31$~meV/atom, found by DFT simulations at $T=0$~K  is reduced by $0.02$~meV/atom if going to the double-layer ordering ($0.29$~meV/atom) the orthorhombic phase still is a metastable state} While the d$_{\perp}$ ordering is insignificantly more favorable than d$_{||}$ in DFT simulations at zero Kelvin, the polarization of both the single and double-layers of \BTO realizes the tetragonal d$_{\perp}$ phase during the heating simulations. 

\ In order to test the stability of the 2-d$_{\perp}$ state with polarization pointing along $z$ we apply a field along x, i.e., perpendicular to the Sr/Ba stacking favoring the 2-d$_{||}$, state see Fig~\ref{fig:PE_DB_TC}. At an electric field of $200$~kV/cm and $400$~kV/cm, the polarization rotates to the field direction, i.e., 2-d$_{\perp}$ and reaches $P_x=7.5$ and $9.7$~$\mu$C/cm$^2$ for 2-d$_{||}$ and double-layer, respectively. However, in the MD simulations this 2-d$_{||}$ is unstable and during field removal, the polarization switches back to $P_z$ at $80$~kV/cm. This results in a double-hysteresis of P(E) with moderate switching fields and narrow hysteresis from $80-200$ and $280-400$~kV/cm for the 2-d$_{||}$ and double-layer, correspondingly. Note that the switching fields are most likely overestimated by the used potentials. 

\section{Summary and Conclusion}

\ We used DFT and molecular dynamics simulations with core-shell potentials to study the impact of Sr-concentration and ordering on the ferroelectric properties of (Ba,Sr)TiO$_3$. With increasing Sr concentration the ferroelectric instability and polarization decrease systematically while the polarizability increases. Importantly due to the different Sr/Ba-O repulsion, the polarization is sensitive to the Sr ordering. Dependent on the local Sr environment deviations of polarization of up to $17$\%, local monoclinic distortions, and large changes in phase stability and transition temperatures exist. Given that all these orderings are close in energy they may locally affect the functional properties such as the polarization switching even for compositionally homogenous materials. The largest effect has to be expected for structured samples with columnar or layered ordering. In this case, particular polarization directions are slightly favored, allowing for a reversible polarization rotation, i.e. ferroelectric double hysteresis for moderate field strengths. Note that this mechanism is different to the double hysteresis observed for SrTiO$_3$ ferroelectric superlattices with larger periodicity, which is driven by depolarization.\cite{aramberri_ferroelectricparaelectric_2022}

\section{Acknowledgements}

\ We acknowledge funding by the German research foundation (DFG, Emmy-Noether group GR 4271/2) and thank Dr. Ruben Khachaturyan for the fruitful discussions.

\begin{figure*}[h!]
    \centering
    \centerline{
   \subfigure[]{\includegraphics[width=0.5\textwidth,clip,trim=0cm 0cm 0cm 0cm]{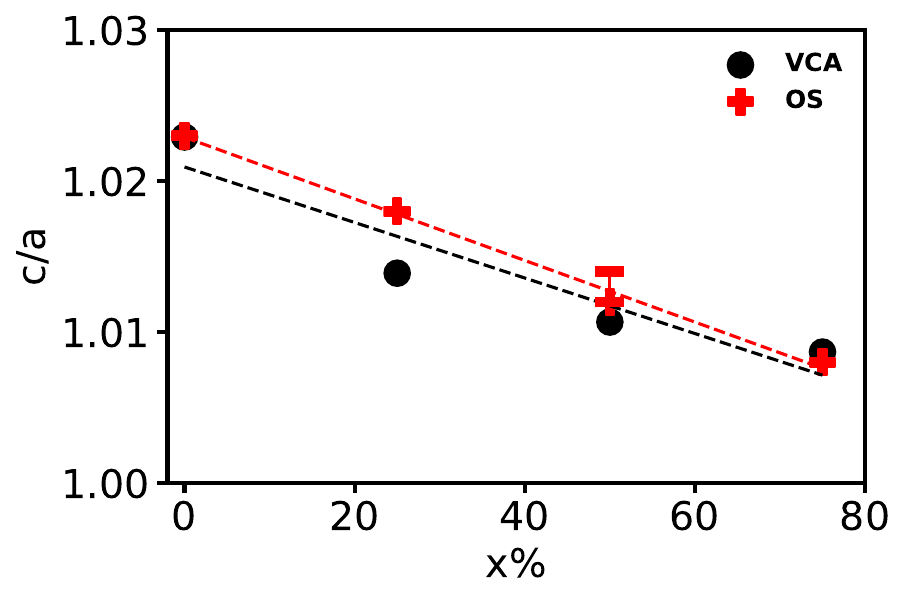}}
  \subfigure[]{\includegraphics[width=0.5\textwidth, clip,trim=0cm 0cm 0cm 0cm]{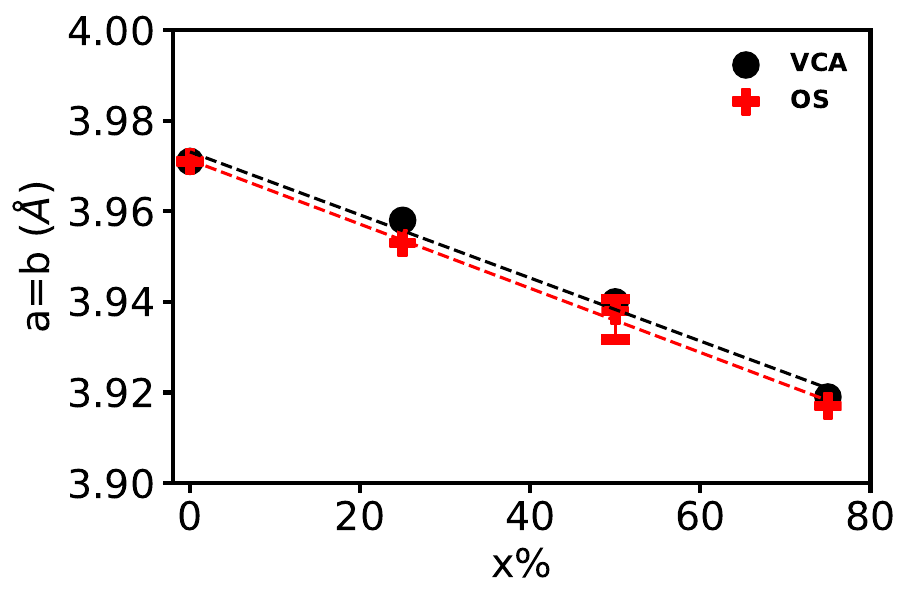}}}
    \centering
    \centerline{
  \subfigure[]{\includegraphics[width=0.5\textwidth, clip,trim=0cm 0cm 0cm 0cm]{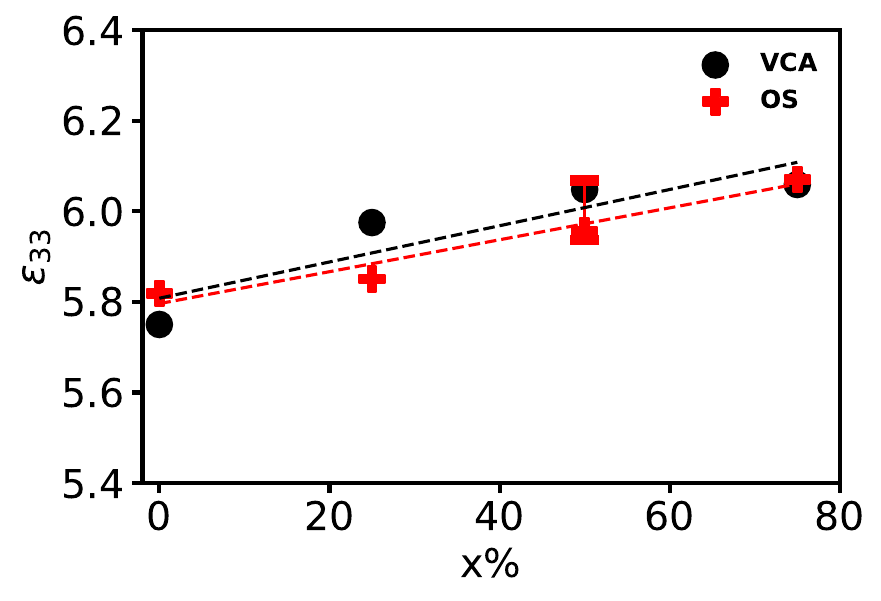}}
  \subfigure[]{\includegraphics[width=0.5\textwidth, clip,trim=0cm 0cm 0cm 0cm]{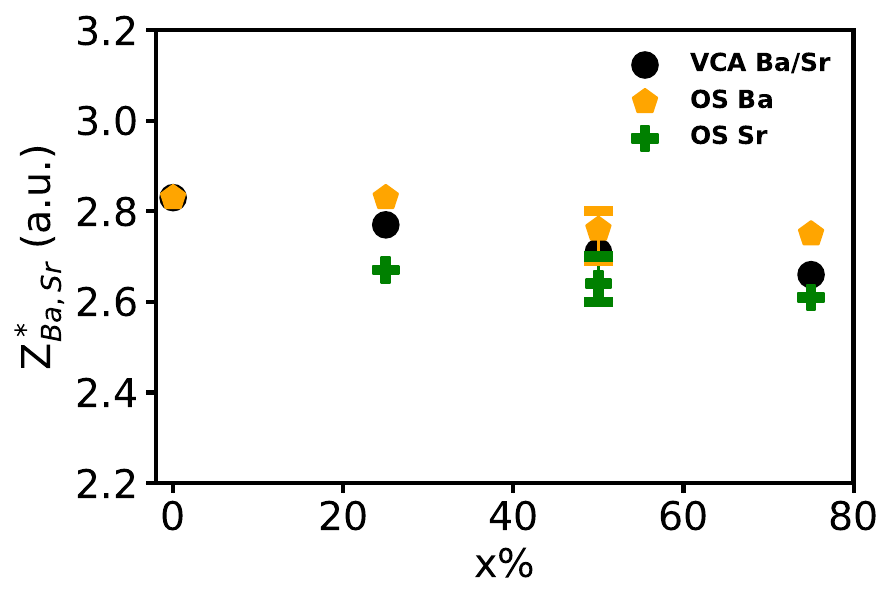}}}
    \caption{Change of properties with varying Sr concentration (x\%), (a) c/a ratio, (b) lattice parameters, (c) high-frequency dielectric constant and (d) Born effective charges of Ba (orange) and Sr (green). The red cross represent the ordered structure (2-d$_{\perp}$)} 
  \label{fig:appendix1}
\end{figure*}

\begin{figure*}[h!]
    \centering
    \centerline{
   \subfigure[]{\includegraphics[width=0.50\textwidth,clip,trim=0cm 0cm 0cm 0cm]{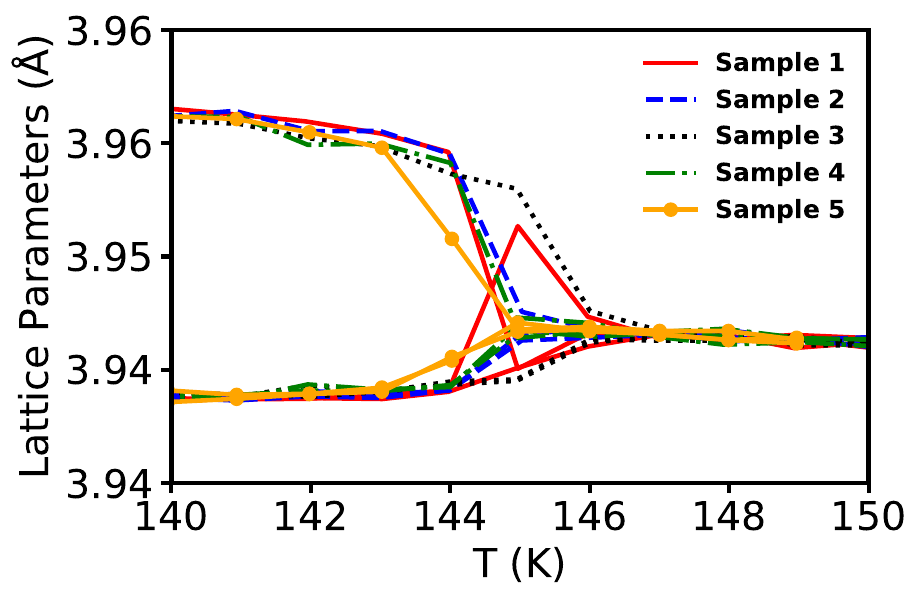}}
  \subfigure[]{\includegraphics[width=0.5\textwidth, clip,trim=0cm 0cm 0cm 0cm]{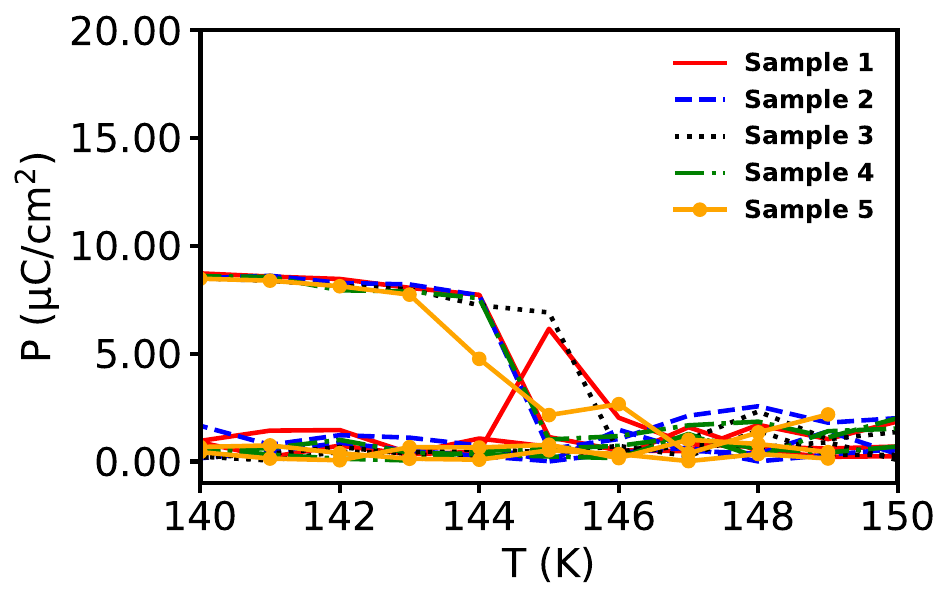}}}
    \caption{Temperature dependency of (a) lattice parameters and (b) polarization components $P_i$ (with i = x, y, z) for different random distributions of Sr in Ba$_{50}$Sr$_{50}$TiO$_3$ in heating simulations. 
    } 
  \label{fig:Samples_stat}

\end{figure*}

\section*{Appendix}

\ In this appendix, we compare lattice parameters,  polarization, and Born effective charges of the pristine materials to literature, see Tab.~\ref{tab:BTO_Pure}. Furthermore, we show the change of lattice parameters, high-frequency dielectric constant, and Z$^{*}_{Ba,Sr}$ with Sr-concentration, see Fig.~\ref{fig:appendix1} and compare the tetragonal to cubic transition for different random orderings of atoms, see Fig.~

\ Moreover, the sampling of the Curie temperature for the randomly distributed Sr atoms within BaTiO$_3$ is presented in Figure~\ref{fig:Samples_stat}.

\begin{table*}[h!]
     \centering
      \caption{Comparison of the calculated lattice constants ($a,b,c$), Born effective charge ($Z^{*}_{33}$), saturation polarization $P_s$ and high frequency dielectric constant (${\epsilon}^{\infty}_{33}$) of cubic {\STO} and {\BTO} as well as of tetragonal {\BTO} to literature.
     \label{tab:BTO_Pure}}
     \begin{tabular}{ccccccccc}
     \hline
 & a=b & c/a & Z$_{Ba}^*$, Z$_{Sr}^*$ & Z$_{Ti}^*$ & Z$_{O_{\perp}}^*$ & Z$_{O_{||}}^*$ & $\epsilon$$_{33}$
         & P$_{s}$ \\
         & (\AA)&&&&&&&($\mu$C/cm$^2$)\\
         \hline
         \BTO (C)  & 3.984~~ & \multirow{3}{*}{1} & 2.74~~ & 7.49~~ & -2.15~~ & -5.93~~ & 6.87~~ & \multirow{3}{*}{--}\\
              & 4.012\cite{kay_xcv_1949}  &  & 2.74\cite{ghosez_lattice_1999} & 7.32\cite{ghosez_lattice_1999}  & -2.14\cite{ghosez_lattice_1999}  & -5.78\cite{ghosez_lattice_1999}& 6.75\cite{ghosez_dynamical_1998} & \\
              &4.000\cite{wang_lattice_2010}&&2.74\cite{evarestov_first-principles_2012}&7.3~~\cite{evarestov_first-principles_2012}&-2.12\cite{evarestov_first-principles_2012}& -5.80\cite{evarestov_first-principles_2012} \\
     
       \BTO (T) & 3.971~~ & 1.023~~ & 2.83~~ & 5.75~~ & -1.95~~ & -4.69~~ &5.75~~ & 30.9~~\\
          & 3.99~\cite{kwei_structures_1993} & 1.011\cite{kwei_structures_1993} & 2.83 \cite{ghosez_born_1995}  & 5.81\cite{ghosez_born_1995}  & -1.99\cite{ghosez_born_1995}  & -4.73\cite{ghosez_born_1995} & 5.81\cite{evarestov_first-principles_2012} & 29\cite{wang_lattice_2010}\\
    & 3.943 \cite{wang_lattice_2010} & 1.023\cite{wang_lattice_2010} & 2.82\cite{siraji_improved_2014}& 5.69\cite{siraji_improved_2014} &-1.94\cite{siraji_improved_2014} & -4.64\cite{siraji_improved_2014} & &\\
          {\STO} (C)  & 3.901~~ & \multirow{3}{*}{1~~} & 2.56~~ & 7.43~~ & -2.05~~ & -5.89~~ & 6.46~~ & \multirow{3}{*}{--}\\
              & 3.884\cite{ghosez_dynamical_1998} &  & 2.56\cite{ghosez_dynamical_1998} & 7.26\cite{ghosez_dynamical_1998} & -2.15\cite{ghosez_dynamical_1998} & -5.73\cite{ghosez_dynamical_1998} & 6.63\cite{lasota_ab_1997} & \\  
                & 3.905\cite{rusevich_electromechanical_2017} &  & 2.54\cite{zhong_giant_1994} & 7.12\cite{zhong_giant_1994} &  -2.00\cite{zhong_giant_1994} &  -5.66\cite{zhong_giant_1994} & & \\  
    \hline
     \end{tabular}
 \end{table*}

\bibliographystyle{unsrt}
\bibliography{Ref.bib}
\end{document}